\documentclass[nonacm,sigconf,fleqn,svgnames,screen]{acmart}

\AtBeginDocument{%
  \providecommand\BibTeX{{%
    \normalfont B\kern-0.5em{\scshape i\kern-0.25em b}\kern-0.8em\TeX}}}

\setcopyright{none}
\copyrightyear{none}
\acmYear{none}
\acmDOI{none}

\definecolor{ml2rblu}{rgb}{0.02,0.27,0.45}
\definecolor{ml2ryel}{rgb}{0.98,0.72,0.18}
\definecolor{ml2rgrn}{rgb}{0.50,0.71,0.18}
\definecolor{ml2rtrq}{rgb}{0.00,0.57,0.57}

\usepackage{amsthm}
\usepackage{mathtools}
\usepackage{bm}
\usepackage[font=footnotesize]{subfig}
\usepackage{booktabs}
\usepackage{array}
\usepackage{enumitem}
\usepackage{fixltx2e}
\usepackage{nicefrac}
\usepackage[capitalize]{cleveref}

\usepackage{wasysym}
\usepackage{xspace}

\usepackage{tikz}
\usetikzlibrary{shapes.geometric, positioning, calc}

\usepackage{algorithm}
\usepackage[noend]{algpseudocode}

\usepackage{hyperref}
\hypersetup{%
  colorlinks=true,
  urlcolor=ml2rtrq,
  linkcolor=ml2rtrq,
  citecolor=ml2rtrq,
  bookmarks=false}

\usepackage{fancyvrb} 
\usepackage{listings}
\lstdefinestyle{pythonstyle}{%
  language=Python,
  tabsize=4,
  backgroundcolor=\color{Gray!10},
  basicstyle=\ttfamily\scriptsize,
  stringstyle=\color{ForestGreen},
  keywordstyle=\color{BlueViolet},
  commentstyle=\itshape\color{DarkRed!90},
  identifierstyle=,
  emphstyle=\color{Blue},
  frame=lines,	
  showstringspaces=false,
  morekeywords={range, len, self, other, lambda, from, import, as, False, True, 
  enumerate, xrange, map, list, set, float, int, min, max, sorted, None},
  fancyvrb=true,
}
\lstnewenvironment{python}[1][]{\lstset{style=pythonstyle,#1}}{}

\lstdefinestyle{cstyle}{%
  language=C,
  tabsize=4,
  backgroundcolor=\color{Gray!10},
  basicstyle=\ttfamily\scriptsize,
  stringstyle=\color{ForestGreen},
  keywordstyle=\color{BlueViolet},
  commentstyle=\itshape\color{DarkRed!90},
  identifierstyle=,
  emphstyle=\color{Blue},
  frame=lines,	
  showstringspaces=false,
  morekeywords={},
  fancyvrb=true,
}
\lstnewenvironment{ccode}[1][]{\lstset{style=cstyle,#1}}{}

\lstdefinestyle{pythonstyletxt}{%
  language=Python,
  tabsize=4,
  basicstyle=\ttfamily\small,
  stringstyle=\color{ForestGreen},
  keywordstyle=\color{BlueViolet},
  commentstyle=\itshape\color{DarkRed!90},
  identifierstyle=,
  emphstyle=\color{Blue},
  xleftmargin=1em,
  showstringspaces=false,
  morekeywords={range, len, self, other, lambda, from, import, as, False, True, 
  enumerate, xrange, map, list, set, float, int, min, max, sorted, with, None},
}
\lstnewenvironment{pythontxt}[1][]{\lstset{style=pythonstyletxt,#1}}{}

\lstdefinestyle{pythonstyletxtsmall}{%
  language=Python,
  tabsize=4,
  basicstyle=\ttfamily\scriptsize,
  stringstyle=\color{ForestGreen},
  keywordstyle=\color{BlueViolet},
  commentstyle=\itshape\color{DarkRed!90},
  identifierstyle=,
  emphstyle=\color{Blue},
  xleftmargin=1em,
  showstringspaces=false,
  morekeywords={range, len, self, other, lambda, from, import, as, False, True, 
  enumerate, xrange, map, list, set, float, int, min, max, sorted, None},
}
\lstnewenvironment{pythontxtsmall}[1][]{\lstset{style=pythonstyletxtsmall,#1}}{}

\newcommand{\putORCID}[1]{
\authornote{\href{https://orcid.org/#1}{\includegraphics[width=2ex]{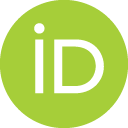}} \href{https://orcid.org/#1}{#1}}
\orcid{#1}}

\usepackage{tikz}
\usetikzlibrary{shapes.geometric, positioning}

\newcommand{\Np}{\emph{NumPy}\xspace}

\newcommand{\QUBO}{\textsf{QUBO}\xspace}
\newcommand{\bQ}{\bm{Q}}
\newcommand{\bx}{\bm{x}}

\newcommand{\tbx}{\tilde{\bm{x}}}
\newcommand{\tx}{\tilde{x}}

\newcommand*{\defeq}{\mathrel{\vcenter{\baselineskip0.5ex\lineskiplimit0pt\hbox{\scriptsize.}\hbox{\scriptsize.}}}%
	=}

\DeclareMathOperator{\CTZ}{CTZ}

\newcommand{\symbspace}[4]{\node (#1) at (#2,#3) {#4\strut}; \draw ([yshift=2pt] $(#1.south west)+(2pt,2pt)$) -- ++(0pt, -2pt) -- ($(#1.south east)+(-2pt,2pt)$) -- ++(0pt, 2pt);}

\begin{document}

\hypersetup{%
  pdftitle={Faster QUBO Brute-Force Solving using Gray Code},
  pdfauthor={Sascha M\"ucke},
  pdfsubject={optimization, scientific python, numpy},
  pdfkeywords={qubo, binary optimization, gray code}
}

\title[Faster QUBO Brute-Force Solving]{Faster QUBO Brute-Force Solving using Gray Code}

\author[S. M\"ucke]{Sascha M\"ucke}
\putORCID{0000-0001-8332-6169}
\affiliation{
  \institution{Lamarr Institute\\ TU Dortmund University}
  \city{Dortmund}
  \state{Germany}
}

\begin{abstract}
This article describes an improved brute-force solving strategy for Quadratic Unconstrained Binary Optimization (\QUBO) problems that is faster than naive approaches and easily parallelizable.
It exploits the Gray code ordering of natural numbers to allow for a more efficient evaluation of the \QUBO objective function.
The implementation in Python is discussed in detail, and an additional C implementation is provided.
\end{abstract}

\maketitle

\section{Introduction}
\emph{Quadratic Unconstrained Binary Optimization} (\QUBO) is the problem of finding a binary vector $\bx^*\in\lbrace 0,1\rbrace^n$ that minimizes the function \begin{equation}\label{eq:qubo}
    f_{\bQ}(\bx)\defeq \bx^{\top}\bQ\bx=\sum_{\substack{i,j\in [n]\\ i\leq j}}Q_{ij}x_ix_j\;,
\end{equation}%
where $\bQ\in\mathbb{R}^{n\times n}$ is an upper triangular matrix, and $[n]$ denotes the set $\lbrace 1,\dots,n\rbrace$.
It is an \textsf{NP}-hard optimization problem \cite{pardalos1992} with numerous applications, ranging from economics \cite{laughhunn1970,hammer1971} over satisfiability \cite{kochenberger2005} and resource allocation \cite{neukart2017,stollenwerk2019} to Machine Learning \cite{bauckhage2018,muecke2019,bauckhage2020,date2020,bauckhage2021}, among others.
In recent years it has gained renewed attention because it is equivalent to the Ising model, which can be solved physically through quantum annealing \cite{kadowaki1998,farhi2000}, for which specialized quantum computers have been developed \cite{d-wavesystems2021}.

Aside from quantum annealing, \QUBO can be solved---exactly or approximately---with a wide range of optimization strategies.
A comprehensive list of approaches can be found in \cite{kochenberger2014}.

The most straightforward way to solve \QUBO is using brute force.
For large $n$, this strategy quickly becomes infeasible, as the number of binary vectors grows exponentially in $n$.
However, for $n$ up to about $30$, which is roughly the order of magnitude currently solvable (approximately) with the hybrid quantum algorithm QAOA \cite{farhi2014}, brute force is an easy-to-implement and reliable way to obtain the minimal bit vector.

In this article we describe a technique to reduce the computational cost of brute-force \QUBO solving by a factor of roughly $n$ compared to the naive approach.
To this end we use the concept of Gray codes to traverse the space $\lbrace 0,1\rbrace^n$ in a way that allows for the value of $f_{\bQ}$ to be updated incrementally, without evaluating all $n\cdot(n+1)/2$ entries of $\bQ$.
A Python implementation using NumPy is developed in \cref{sec:practice}, and ways to improve running time by using just-in-time compilation and parallelization are described.

\section{Theory}
Generally, computing the value of $f_{\bQ}(\bx)$ requires evaluating a sum over all entries of $\bQ$.
However, if the value of $\bx$ is already known to be $v$, and we want to compute the value of $\tbx$ that differs from $\bx$ in only a single bit, we can ``update'' $v$ to obtain $f_{\bQ}(\tbx)$, which requires only $n$ values of $\bQ$ (see \cref{sec:updating}).

Now all that is missing is a way to traverse all vectors in $\lbrace 0,1\rbrace^n$ by changing only one bit at a time.
Luckily, we can take inspiration from Gray codes, which have exactly the right properties for this task and are easy to compute (see \cref{sec:gray}).

\subsection{Updating QUBO values}\label{sec:updating}

Let $\bQ\in\lbrace 0,1\rbrace^{n\times n}$ be a fixed \QUBO matrix, and $\bx\in\lbrace 0,1\rbrace^n$ a binary vector.
Further, let $v=f_{\bQ}(\bx)$.
Now, assume we want to flip the $\ell$-th bit of $\bx$ to obtain $\tbx$ and calculate the new value $\tilde{v}=f_{\bQ}(\tbx)$.
Instead of calculating $\tbx^{\top}\bQ\tbx$ explicitly, we can look at the difference between $v$ and $\tilde{v}$: \begin{align}
    \tilde{v} &= v+\Delta_{\ell} \nonumber\\
    \Leftrightarrow ~\Delta_{\ell} &= \tilde{v} - v \nonumber\\
        &= \sum_{i\leq j} Q_{ij}(\tx_i\tx_j-x_ix_j) \nonumber\\
        &= s_{\ell}\biggl(\sum_{i=1}^{\ell-1}Q_{i\ell}x_i + Q_{\ell\ell} + \sum_{j=\ell+1}^nQ_{\ell j}x_j\biggr)\;, \label{eq:delta}
\end{align}%
where $s_\ell=\tx_\ell-x_\ell\in\lbrace -1,+1\rbrace$.

As we can see from the last line, we only need to read the elements in the $\ell$-th row and column of $\bQ$ to calculate $\Delta_\ell$.
This reduces the computational cost from $\mathcal{O}(n^2)$ to $\mathcal{O}(n)$ per evaluation of $f_{\bQ}$.
As the vector $\bm{0}\defeq (0,\dots,0)^{\top}$ always has value $0$ for any $\bQ$, we can start there and never need to fully calculate \cref{eq:qubo}, if we find a way to successively flip bits and traverse all $\bx\in\lbrace 0,1\rbrace^n$.
This problem is addressed next.

\subsection{Gray Code}\label{sec:gray}

\begin{figure}[t]
    \centering
    \includegraphics[width=.5\columnwidth]{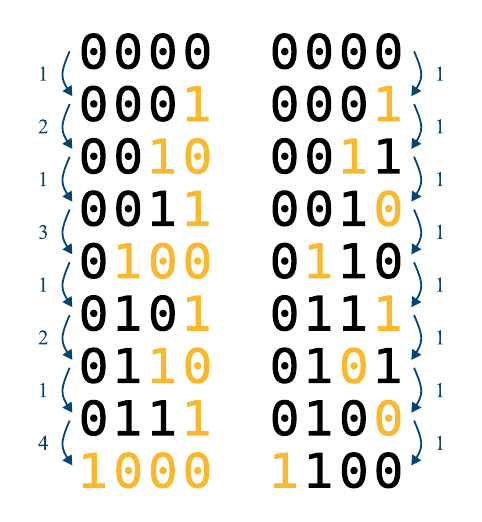}
    \caption{Gray Code}
    \label{fig:graycode}
\end{figure}

When starting at 0 and counting up, the respective binary representations of successive numbers often differ in more than one bit (see \cref{fig:graycode}, left).
For example, to get from $2^k-1$ to $2^k$ for any $k\geq 0$, exactly $k+1$ bits must be flipped.
These transitions where many bits flip at once are nown as \emph{Hamming cliffs} \cite{caruana1988}.
Therefore, this way of traversing all bitvectors cannot be used with the aforementioned updating method.

\emph{Gray code} is an ordering of the natural numbers $\pi:\mathbb{N}_0\rightarrow\mathbb{N}_0$ that removes Hamming cliffs, i.e., for any $k\in\mathbb{N}_0$ we find that $\pi(k)$ and $\pi(k+1)$ differ in one bit when represented in binary (see \cref{fig:graycode}, right).
The sequence $\pi(0),\dots,\pi(2^k-1)$ for any $k>1$ can be constructed recursively in binary from the sequence for $k-1$ through \begin{equation}
    \pi(\ell) \defeq \begin{cases}
        \mathtt{0}\cdot \pi(\ell) &\text{if } \ell<2^{k-1} \\
        \mathtt{1}\cdot \pi(2^k-\ell-1) &\text{else}        
    \end{cases}
\end{equation}%
Here, $\cdot$ denotes string concatenation.
For $k=1$ the code is just $(\mathtt{0}, \mathtt{1})$.

As only one bit is flipped at a time, we can express $\pi$ equivalently as the sequence of bit indices to flip, starting at $\bm{0}$ and using 0-indexing.
This sequence turns out to be $0,1,0,2,0,1,0,3,0,1,0,2,0,\dots$, which is known as the \emph{binary carry sequence}\footnote{\url{https://oeis.org/A007814}}.
Interestingly, it coincides with the number of trailing zeros when writing the natural numbers $1,2,3,\dots$ in binary, which gives us a very efficient way to calculate its terms on a CPU (see \cref{sec:practice}).
The number of trailing zeros of a number $k$ is denoted by $\CTZ(k)$.

\begin{algorithm}[t]
\caption{Improved \QUBO Brute-Force Solving}\label{alg:bruteforce}
\begin{algorithmic}
\State $\bx\gets\bm{0}$
\State $v\gets 0$
\State $v^*\gets \infty$
\State $i\gets 1$
\While{$i < 2^n$}
    \State $\ell\gets\CTZ(i)$
    \State $x_{\ell}\gets 1-x_{\ell}$
    \State $\Delta_{\ell}\gets\sum_{i=1}^{\ell-1}Q_{i\ell}x_i + Q_{\ell\ell} + \sum_{j=\ell+1}^nQ_{\ell j}x_j$
    \State $v\gets v+(2x_{\ell}-1)\Delta_{\ell}$
    \If{$v<v^*$}
        \State $\bx^*\gets\bx$ \Comment{memorize minimizing $\bx$\dots}
        \State $v^*\gets v$ \Comment{\dots and its value}
    \EndIf
    \State $i\gets i+1$
\EndWhile
\Return $\bx^*, ~v^*$
\end{algorithmic}
\end{algorithm}

Now that we have all necessary ingredients for a more efficient brute-force strategy, we can combine everything into \cref{alg:bruteforce}.
In the following section, we show how to put it into practice by implementing it in Python, using \Np and \emph{Numba} to improve the running time, which is why we need to \begin{pythontxt}
import numpy as np
from numba import njit
\end{pythontxt}

\section{Practice}\label{sec:practice}

For comparison, two versions of brute-force \QUBO solving methods will be implemented here:
A naive approach and the improved version described in \cref{alg:bruteforce}.

Assume that the parameters of the \QUBO problem instance are given as an upper-triangular \Np matrix \verb+Q+ of shape \verb+(n, n)+.
For the naive approach, we will simply loop over all binary vectors in ``ascending order'', i.e., we count up from 0 through $2^n-1$ and convert each number to its binary representation.
This can be done somewhat efficiently by using NumPy's broadcasting functionality:
Assume we want to represent an integer \verb+k+ as a binary vector of length \verb+n+.
We can write: \begin{pythontxt}
places = 2**np.arange(n)
x = (k & places) > 0
\end{pythontxt}

The operator \verb+&+ performs an element-wise logical AND on the binary representation of \verb+k+ and each element in \verb+k+.
If the $i$-th bit of \verb+k+ is set, \verb+k & 2**i+ equals \verb+2**i+, otherwise it evaluates to 0.
Thus, to make the result binary, we can simply check if it is greater than zero.
Doing this for every bit index $i$ and collecting the result in an array yields the binary representation of \verb+k+.

To evaluate the \QUBO objective function \cref{eq:qubo}, we can simply perform the twofold matrix-vector product as \begin{pythontxt}
    v = x @ Q @ x
\end{pythontxt}

All that is missing is the loop over all $2^n$ vectors, and a running minimality check.
In summary, the naive brute-force algorithm could look as follows:
\begin{python}
def naive_brute_force(Q):
    n = Q.shape[0]
    # initialize bit vector and value
    x = np.zeros(n)
    v = 0
    # initialize minimal bit vector and value
    x_min = np.zeros(n)
    v_min = 0

    places = 2 ** np.arange(n) # can be outside loop
    for k in range(1, 2**n):
        x[:] = (k & places) > 0 # get binary vector from k
        v = x @ Q @ x    # get QUBO objective value
        if v < v_min:    # check for minimality
            x_min[:] = x # memorize binary vector..
            v_min = v    # ..and value
    return x_min, v_min
\end{python}
\noindent Notice that we can start the loop at 1 and directly set \verb+x_min+ and \verb+v_min+ to 0, as $f_{\bQ}(\bm{0})=0$ regardless of $\bQ$.

Now let us implement the improved version described in \cref{alg:bruteforce}.
There are only few changes compared with the naive version: \begin{enumerate}
    \item Update $\bx$ incrementally in Gray code order
    \item Update function value $v$ incrementally.
\end{enumerate}
To address the first point, we need to implement the $\CTZ$ function.
In Python, we have a builtin method \verb+int.bit_count()+, which counts the 1 bits in the binary representation of an integer.
We can use this function to determine $\CTZ$ of a number \verb+k+ by \begin{pythontxt}
(k ^ (k-1)).bit_count()-1
\end{pythontxt}
Intuitively, if a number has $\ell$ trailing zeros, then subtracting 1 yields a number with $\ell$ trailing ones and a zero in $(\ell+1)$-th place.
The XOR (\verb+^+) of \verb+k+ and \verb+k-1+ thus consists of $\ell+1$ ones, from which we need to subtract 1 to get just $\ell$.
This is the index we need to flip $\bx$ at: \begin{pythontxt}
x[l] = 1-x[l]
\end{pythontxt}

The second point can be implemented by first making the (triangular) \QUBO matrix symmetric, which lets us simply read one row instead of both row and column.
Further, we save the diagonal containing the linear terms in a separate array: \begin{pythontxt}
qua  = np.triu(Q, 1) # clear diagonal
qua += qua.T         # make symmetric
lin  = np.diag(Q)
\end{pythontxt}

\noindent This lets us write \cref{eq:delta} as \begin{pythontxt}
delta = (2*x[l]-1) * (qua[l]@x + lin[l])
\end{pythontxt}
\noindent and the complete implementation reads as follows:
\begin{python}
def improved_brute_force(Q):
    n = Q.shape[0]
    # initialize bit vector and value
    x = np.zeros(n)
    v = 0
    # initialize minimal bit vector and value
    x_min = np.zeros(n)
    v_min = 0
    # separate Q
    qua  = np.triu(Q, 1)
    qua += qua.T
    lin  = np.diag(Q)
    for k in range(1, 2**n):
        l = (k ^ (k-1)).bit_count()-1
        x[l] = 1-x[l]
        delta = (2*x[l]-1) * (qua[l]@x + lin[l])
        v += delta
        if v < v_min:
            x_min[:] = x
            v_min = v
    return x_min, v_min
\end{python}

While this code works perfectly fine, Python loops are comparatively slow.
For this reason, we can use the package \verb+numba+, which allows for just-in-time (JIT) compilition of certain, structurally simple functions to C.
In many cases, this increases the performance considerably.
Most NumPy functionality is covered by \verb+numba+.
We only need to change a single line of our code, as the Python function \verb+int.bit_count()+ is not supported.
We can replace the respective line with \begin{pythontxt}
l = int(np.log2(k ^ (k-1)))
\end{pythontxt}%
which is functionally equivalent.
To enable JIT compilation, we simply need to add a decorator before both function definitions: \begin{pythontxt}
from numba import njit

@njit
def naive_brute_force(Q):
    ...

@njit
def improved_brute_force(Q):
    ...
\end{pythontxt}

\begin{figure}[t]
    \centering
    \includegraphics[width=\columnwidth]{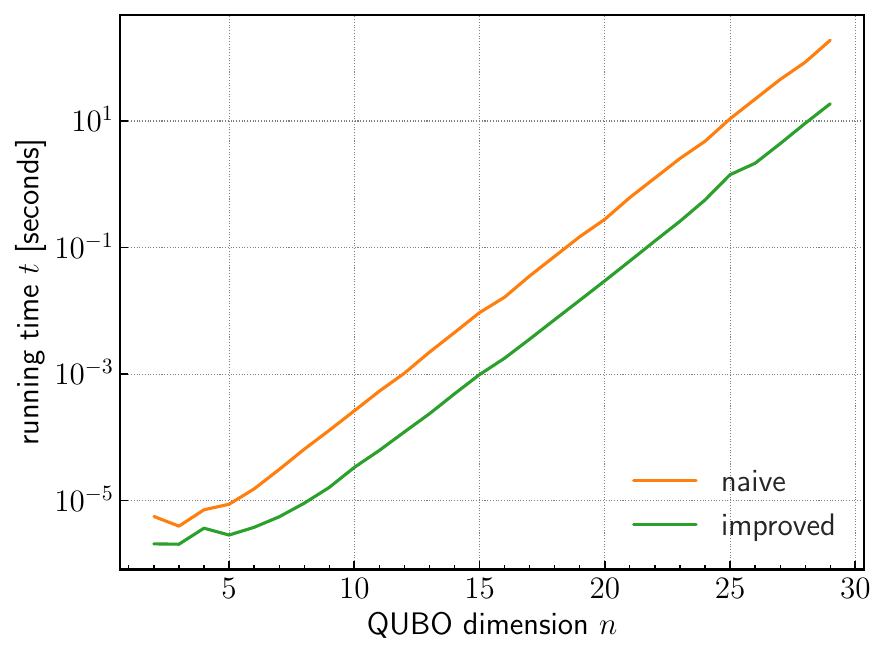}
    \caption{Running time comparison between naive brute-force solving and \cref{alg:bruteforce}; lower is better.}
    \label{fig:timing}
\end{figure}

The resulting running times of both methods are shown in \cref{fig:timing}.
The experiment was conducted on an Intel Core i7-8700 CPU running Python 3.10.5 on an Arch-based Linux system.
Running times are averaged over 10 random \QUBO instances.
We can clearly see that the improved version is, on average, faster by a factor of about 10 than the naive version on the tested value range.

\subsection{Parallelization}\label{sec:parallel}

The code presented in the previous section can be easily modified to be executed in parallel.
To this end, we can fix the last $m$ bits of every $\bx$ and let the brute-force loop only run over the first $n-m$ bits, and their results combined by taking the minimum.

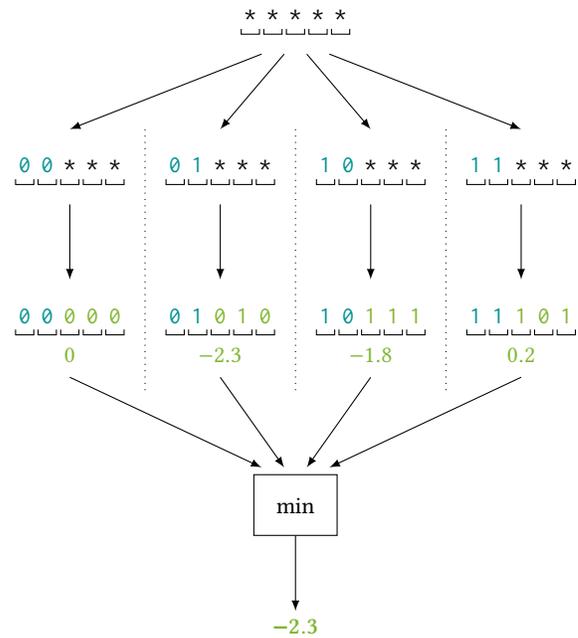
\begin{figure}
\centering
\begin{tikzpicture}
    \pgfmathsetmacro{\gap}{0.3}
    \foreach \x in {0,...,4} {
        \symbspace{a0}{3+\x*\gap}{4}{\texttt{*}}}
    \foreach \z in {0,...,3} {
        \draw[-latex] (3.15+\z*\gap, 3.5) -- (2*\gap+2*\z, 2.5);}
    \foreach \xa in {0,1} { \foreach \xb in {0,1} {
        \pgfmathsetmacro{\x}{\xa*2+\xb} 
        \symbspace{b00}{\x*2}{2}{\color{ml2rtrq}\texttt{\xa}}
        \symbspace{b01}{\x*2+\gap}{2}{\color{ml2rtrq}\texttt{\xb}}
        \foreach \z in {0,1,2} {
            \symbspace{b0}{\x*2+2*\gap+\z*\gap}{2}{\texttt{*}}}}}
    \foreach \z in {0,...,3} {
        \draw[-latex] (2*\gap+2*\z, 1.5) -- ++(0, -1);}
    \foreach \xa in {0,1} { \foreach \xb in {0,1} {
        \pgfmathsetmacro{\x}{\xa*2+\xb} 
        \symbspace{c00}{\x*2}{0}{\color{ml2rtrq}\texttt{\xa}}
        \symbspace{c01}{\x*2+\gap}{0}{\color{ml2rtrq}\texttt{\xb}}
        \foreach \z in {0,1,2} {
            \pgfmathsetmacro{\v}{int(Mod(\xa+\xb*\z,2))}
            \symbspace{c0}{\x*2+2*\gap+\z*\gap}{0}{\color{ml2rgrn}\texttt{\v}}}}}
    \node (v0) at (2*\gap, -0.5) {\color{ml2rgrn}$0$};
    \node (v0) at (2+2*\gap, -0.5) {\color{ml2rgrn}$-2.3$};
    \node (v0) at (4+2*\gap, -0.5) {\color{ml2rgrn}$-1.8$};
    \node (v0) at (6+2*\gap, -0.5) {\color{ml2rgrn}$0.2$};
    \node[draw, inner sep=3mm] (min) at (3+2*\gap, -2.5) {$\min$};
    \foreach \z in {0,...,3} {
        \draw[-latex] (2*\gap+2*\z, -0.8) -- (3.15+\z*\gap, -2);}
    \draw[-latex] (min.south) -- ++(0, -1);
    \node at ($(min.south)+(0,-1.2)$) {\color{ml2rgrn}$\bm{-2.3}$};
    \foreach \x in {1.6, 3.6, 5.6} {
        \draw[dotted] (\x, 2.5) -- (\x, -1);}
\end{tikzpicture}
\caption{Parallel brute-force solving by fixing $2$ bits of a $5$ bit vector.
The bits marked with \texttt{*} are to be optimized.
The steps seperated by dotted lines can be performed on four different CPUs in parallel.}
\label{fig:subspaces}
\end{figure}

This way, we obtain $2^m$ sub-problems, one for each assignment of $m$ bits, which can be evaluated in parallel and the results combined by taking their minimum.
This is exemplarily shown in \cref{fig:subspaces} for $n=5$ and $m=2$.
Using this technique, if there are $P$ CPUs available, the number of additional bits that can be solved in the same amount of time is $\lfloor\log_2(P)\rfloor$.

A C implementation of \cref{alg:bruteforce} using this parallelization method is given in \cref{sec:cimpl}.

\section{Conclusion}

In this work we have seen that brute-force solving \QUBO problems can be made more efficient by updating the function value incrementally.
To this end, we can leverage Gray codes to obtain an order of traversing all $n$-bit vectors in a way that allows us to compute said function value update more efficiently, reducing the computational complexity within the loop from $\mathcal{O}(n^2)$ to $\mathcal{O}(n)$.
Moreover, we have seen how to implement the resulting algorithm in Python.
We have further seen how \verb+numba+ can improve performance through JIT compilation, which requires next to no changes to the code.
Finally, a parallelization scheme was presented that is very easy to implement, as demonstrated with the C code given in \cref{sec:cimpl}.

While brute-force solving is still infeasible for large $n$, having a fast implementation of such an algorithm is still a valuable tool for research purposes.
Often, experiments are conducted on low-dimensional \QUBO instances, e.g., to validate theoretical properties.
Current quantum algorithms like QAOA can solve \QUBO instances with a low two-digit number of qubits.
On our machine we are able to brute-force \QUBO instances with $n=30$ in under 4 seconds, which is valuable for quickly obtaining ground-truth solutions for benchmark problems.
Lastly, the algorithm presented here uses some interesting techniques to improve efficiency, that may serve as inspiration for similar problems.

The multi-threaded C implementation of this algorithm will be included in the upcoming version \dots of the Python package \dots (\textit{left out for anonymous version}).
The multi-threaded C implementation of \cref{alg:bruteforce} is included in the the latest version of my Python package \verb+qubolite+\footnote{\url{https://github.com/smuecke/qubolite}}: \begin{pythontxt}
    pip install qubolite
\end{pythontxt}

\section*{Acknowledgments} 

This research has been funded by the Federal Ministry of Education and Research of Germany and the state of North-Rhine Westphalia as part of the Lamarr Institute for Machine Learning and Artificial Intelligence.

\appendix
\section{C Implementation}\label{sec:cimpl}

The following code is an implementation of \cref{alg:bruteforce} in C, which additionally uses the parallelization scheme described in \cref{sec:parallel}.
For multi-threading, the \emph{OpenMP} interface is used (be sure to use the \verb+-fopenmp+ flag when compiling with \verb+gcc+ on Linux).
The method \verb+brute_force_parallel+ expects the \QUBO matrix as a 2D array of double-precision floats.

\begin{lstlisting}[style=cstyle]
#include <stdio.h>
#include <stdint.h>
#include <stdlib.h>
#include <string.h>
#include <omp.h>

typedef unsigned char bit;

struct _result_t {
    bit *x_min;
    double v_min;
} typedef result_t;

result_t brute_force_worker(
	double **qubo,
	const size_t n,
	const size_t m) {
    bit x[n];
    memset(x, 0, n);
    double v = 0;
    bit *x_min = (bit*) malloc(n);
    double v_min;
    // set sub-vector for each thread
    size_t p = omp_get_thread_num();
    for (size_t i=0; i<m; ++i)
        x[n-i-1] = (p & (1<<i))>0 ? 1 : 0;
    // evaluate QUBO once on initial vector
    for (size_t i=n-m; i<n; ++i) {
        if (x[i] <= 0)
            continue;
	v += qubo[i][i];
        for (size_t j=i+1; j<n; ++j)
            v += x[j]*qubo[i][j];
    }
    // set initial minimal values
    memcpy(x_min, x, n);
    v_min = v;

    size_t l;
    double delta;
    for (int64_t i=1; i<(1<<(n-m)); ++i) {
        l = __builtin_ctzll(i);
        x[l] ^= 1; // flip bit
        // calculate function value update
        delta = 0;
        for (size_t j=0; j<l; ++j)
            delta += x[j]*qubo[j][l];
        delta += qubo[l][l];
        for (size_t j=l+1; j<n; ++j)
            delta += x[j]*qubo[l][j];
        v += x[l] ? delta : -delta;
        if (v < v_min) {
            memcpy(x_min, x, n);
            v_min = v;
        }
    }
    result_t result = {x_min, v_min};
    return result;
}

result_t brute_force_parallel(double **qubo, const size_t n) {
    // determine max. number of threads
    const int64_t p_max = omp_get_max_threads();
    // take floor of log2 of p_max to get
    // max number of bits that can be fixed
    const size_t m = 63 - __builtin_clzll(p_max);
    const size_t M = 1<<m; // number of threads
    // allocate space for sub-results
    result_t results[M];
    // make sure to use exactly 2**m threads
    omp_set_dynamic(0);
    #pragma omp parallel num_threads(M)
    {
        results[omp_get_thread_num()]
	    = brute_force_worker(qubo, n, m);
    }
    // get global minimum
    bit *x_min = (bit*) malloc(n);
    double v_min = 0;
    double v;
    for (size_t i=0; i<M; ++i) {
	v = results[i].v_min;
        if (v<v_min) {
	    memcpy(x_min, results[i].x_min, n);
            v_min = v;
	}
	free(results[i].x_min);
    }
    result_t result = {x_min, v_min};
    return result;
}

int main() {
    const size_t n = 8;
    double **qubo = (double**) malloc(n*sizeof(double*));
    for (size_t i=0; i<n; ++i)
	qubo[i] = (double*) malloc(n*sizeof(double));

    for (size_t i=0; i<n; ++i) {
        for (size_t j=0; j<n; ++j) {
            if (i>j)
                qubo[i][j] = 0;
	    else
		qubo[i][j] = (double) (i-j+2);
        }
    }

    result_t result = brute_force_parallel((double**) qubo, n);
    printf("minimum vector: ");
    for (size_t i=0; i<n; ++i)
	printf("%d", result.x_min[i]);
    printf("\nminimum value:  %f\n", result.v_min);

    // free all allocated resources
    free(result.x_min);
    for (size_t i=0; i<n; ++i)
	free(qubo[i]);
    free(qubo);
}
\end{lstlisting}


\begin{thebibliography}{17}
	
	\ifx \showCODEN    \undefined \def \showCODEN     #1{\unskip}     \fi
	\ifx \showDOI      \undefined \def \showDOI       #1{#1}\fi
	\ifx \showISBNx    \undefined \def \showISBNx     #1{\unskip}     \fi
	\ifx \showISBNxiii \undefined \def \showISBNxiii  #1{\unskip}     \fi
	\ifx \showISSN     \undefined \def \showISSN      #1{\unskip}     \fi
	\ifx \showLCCN     \undefined \def \showLCCN      #1{\unskip}     \fi
	\ifx \shownote     \undefined \def \shownote      #1{#1}          \fi
	\ifx \showarticletitle \undefined \def \showarticletitle #1{#1}   \fi
	\ifx \showURL      \undefined \def \showURL       {\relax}        \fi
	\providecommand\bibfield[2]{#2}
	\providecommand\bibinfo[2]{#2}
	\providecommand\natexlab[1]{#1}
	\providecommand\showeprint[2][]{arXiv:#2}
	
	\bibitem[\protect\citeauthoryear{Bauckhage, Beaumont, and M{\"u}ller}{Bauckhage
		et~al\mbox{.}}{2021}]%
	{bauckhage2021}
	\bibfield{author}{\bibinfo{person}{Christian Bauckhage},
		\bibinfo{person}{Fabrice Beaumont}, {and} \bibinfo{person}{Sebastian
			M{\"u}ller}.} \bibinfo{year}{2021}\natexlab{}.
	\newblock \bibinfo{booktitle}{\emph{{{ML2R Coding Nuggets}}: {{Hopfield Nets}}
			for {{Hard Vector Quantization}}}}.
	\newblock
	
	
	\bibitem[\protect\citeauthoryear{Bauckhage, Ojeda, Sifa, and Wrobel}{Bauckhage
		et~al\mbox{.}}{2018}]%
	{bauckhage2018}
	\bibfield{author}{\bibinfo{person}{Christian Bauckhage}, \bibinfo{person}{Cesar
			Ojeda}, \bibinfo{person}{Rafet Sifa}, {and} \bibinfo{person}{Stefan Wrobel}.}
	\bibinfo{year}{2018}\natexlab{}.
	\newblock \showarticletitle{Adiabatic Quantum Computing for Kernel k= 2 Means
		Clustering.}. In \bibinfo{booktitle}{\emph{{{LWDA}}}}.
	\bibinfo{pages}{21--32}.
	\newblock
	
	
	\bibitem[\protect\citeauthoryear{Bauckhage, Ramamurthy, and Sifa}{Bauckhage
		et~al\mbox{.}}{2020}]%
	{bauckhage2020}
	\bibfield{author}{\bibinfo{person}{C. Bauckhage}, \bibinfo{person}{R.
			Ramamurthy}, {and} \bibinfo{person}{R. Sifa}.}
	\bibinfo{year}{2020}\natexlab{}.
	\newblock \showarticletitle{Hopfield {{Networks}} for {{Vector Quantization}}}.
	In \bibinfo{booktitle}{\emph{Artificial {{Neural Networks}} and {{Machine
					Learning}} \textendash{} {{ICANN}} 2020}} \emph{(\bibinfo{series}{Lecture
			{{Notes}} in {{Computer Science}}})}. \bibinfo{publisher}{{Springer
			International Publishing}}, \bibinfo{pages}{192--203}.
	\newblock
	\urldef\tempurl%
	\url{https://doi.org/10.1007/978-3-030-61616-8_16}
	\showDOI{\tempurl}
	
	
	\bibitem[\protect\citeauthoryear{Caruana and Schaffer}{Caruana and
		Schaffer}{1988}]%
	{caruana1988}
	\bibfield{author}{\bibinfo{person}{Richard~A. Caruana} {and}
		\bibinfo{person}{J.~David Schaffer}.} \bibinfo{year}{1988}\natexlab{}.
	\newblock \showarticletitle{Representation and {{Hidden Bias}}: {{Gray}} vs.
		{{Binary Coding}} for {{Genetic Algorithms}}}.
	\newblock In \bibinfo{booktitle}{\emph{Machine {{Learning Proceedings}} 1988}}.
	\bibinfo{publisher}{{Morgan Kaufmann}}, \bibinfo{pages}{153--161}.
	\newblock
	\urldef\tempurl%
	\url{https://doi.org/10.1016/B978-0-934613-64-4.50021-9}
	\showDOI{\tempurl}
	
	
	\bibitem[\protect\citeauthoryear{{D-Wave Systems}}{{D-Wave Systems}}{2021}]%
	{d-wavesystems2021}
	\bibfield{author}{\bibinfo{person}{{D-Wave Systems}}.}
	\bibinfo{year}{2021}\natexlab{}.
	\newblock \bibinfo{booktitle}{\emph{Technical {{Description}} of the {{D-Wave
					Quantum Processing Unit}}}}.
	\newblock
	
	
	\bibitem[\protect\citeauthoryear{Date, Arthur, and {Pusey-Nazzaro}}{Date
		et~al\mbox{.}}{2020}]%
	{date2020}
	\bibfield{author}{\bibinfo{person}{Prasanna Date}, \bibinfo{person}{Davis
			Arthur}, {and} \bibinfo{person}{Lauren {Pusey-Nazzaro}}.}
	\bibinfo{year}{2020}\natexlab{}.
	\newblock \showarticletitle{{{QUBO Formulations}} for {{Training Machine
				Learning Models}}}.
	\newblock \bibinfo{journal}{\emph{arXiv:2008.02369 [physics, stat]}}
	(\bibinfo{year}{2020}).
	\newblock
	\showeprint[arxiv]{physics, stat/2008.02369}
	
	
	\bibitem[\protect\citeauthoryear{Farhi, Goldstone, and Gutmann}{Farhi
		et~al\mbox{.}}{2014}]%
	{farhi2014}
	\bibfield{author}{\bibinfo{person}{Edward Farhi}, \bibinfo{person}{Jeffrey
			Goldstone}, {and} \bibinfo{person}{Sam Gutmann}.}
	\bibinfo{year}{2014}\natexlab{}.
	\newblock \showarticletitle{A Quantum Approximate Optimization Algorithm}.
	\newblock \bibinfo{journal}{\emph{arXiv preprint arXiv:1411.4028}}
	(\bibinfo{year}{2014}).
	\newblock
	\showeprint[arxiv]{1411.4028}
	
	
	\bibitem[\protect\citeauthoryear{Farhi, Goldstone, Gutmann, and Sipser}{Farhi
		et~al\mbox{.}}{2000}]%
	{farhi2000}
	\bibfield{author}{\bibinfo{person}{Edward Farhi}, \bibinfo{person}{Jeffrey
			Goldstone}, \bibinfo{person}{Sam Gutmann}, {and} \bibinfo{person}{Michael
			Sipser}.} \bibinfo{year}{2000}\natexlab{}.
	\newblock \showarticletitle{Quantum Computation by Adiabatic Evolution}.
	\newblock \bibinfo{journal}{\emph{arXiv preprint quant-ph/0001106}}
	(\bibinfo{year}{2000}).
	\newblock
	\showeprint[arxiv]{quant-ph/0001106}
	
	
	\bibitem[\protect\citeauthoryear{Hammer and Shlifer}{Hammer and
		Shlifer}{1971}]%
	{hammer1971}
	\bibfield{author}{\bibinfo{person}{Peter~L Hammer} {and}
		\bibinfo{person}{Eliezer Shlifer}.} \bibinfo{year}{1971}\natexlab{}.
	\newblock \showarticletitle{Applications of Pseudo-{{Boolean}} Methods to
		Economic Problems}.
	\newblock \bibinfo{journal}{\emph{Theory and decision}} \bibinfo{volume}{1},
	\bibinfo{number}{3} (\bibinfo{year}{1971}), \bibinfo{pages}{296--308}.
	\newblock
	
	
	\bibitem[\protect\citeauthoryear{Kadowaki and Nishimori}{Kadowaki and
		Nishimori}{1998}]%
	{kadowaki1998}
	\bibfield{author}{\bibinfo{person}{Tadashi Kadowaki} {and}
		\bibinfo{person}{Hidetoshi Nishimori}.} \bibinfo{year}{1998}\natexlab{}.
	\newblock \showarticletitle{Quantum Annealing in the Transverse {{Ising}}
		Model}.
	\newblock \bibinfo{journal}{\emph{Physical Review E}} \bibinfo{volume}{58},
	\bibinfo{number}{5} (\bibinfo{year}{1998}), \bibinfo{pages}{5355}.
	\newblock
	
	
	\bibitem[\protect\citeauthoryear{Kochenberger, Glover, Alidaee, and
		Lewis}{Kochenberger et~al\mbox{.}}{2005}]%
	{kochenberger2005}
	\bibfield{author}{\bibinfo{person}{Gary Kochenberger}, \bibinfo{person}{Fred
			Glover}, \bibinfo{person}{Bahram Alidaee}, {and} \bibinfo{person}{Karen
			Lewis}.} \bibinfo{year}{2005}\natexlab{}.
	\newblock \showarticletitle{Using the Unconstrained Quadratic Program to Model
		and Solve {{Max}} 2-{{SAT}} Problems}.
	\newblock \bibinfo{journal}{\emph{International Journal of Operational
			Research}} \bibinfo{volume}{1}, \bibinfo{number}{1-2} (\bibinfo{year}{2005}),
	\bibinfo{pages}{89--100}.
	\newblock
	
	
	\bibitem[\protect\citeauthoryear{Kochenberger, Hao, Glover, Lewis, L{\"u},
		Wang, and Wang}{Kochenberger et~al\mbox{.}}{2014}]%
	{kochenberger2014}
	\bibfield{author}{\bibinfo{person}{Gary Kochenberger}, \bibinfo{person}{Jin-Kao
			Hao}, \bibinfo{person}{Fred Glover}, \bibinfo{person}{Mark Lewis},
		\bibinfo{person}{Zhipeng L{\"u}}, \bibinfo{person}{Haibo Wang}, {and}
		\bibinfo{person}{Yang Wang}.} \bibinfo{year}{2014}\natexlab{}.
	\newblock \showarticletitle{The Unconstrained Binary Quadratic Programming
		Problem: A Survey}.
	\newblock \bibinfo{journal}{\emph{Journal of Combinatorial Optimization}}
	\bibinfo{volume}{28}, \bibinfo{number}{1} (\bibinfo{year}{2014}),
	\bibinfo{pages}{58--81}.
	\newblock
	
	
	\bibitem[\protect\citeauthoryear{Laughhunn}{Laughhunn}{1970}]%
	{laughhunn1970}
	\bibfield{author}{\bibinfo{person}{DJ Laughhunn}.}
	\bibinfo{year}{1970}\natexlab{}.
	\newblock \showarticletitle{Quadratic Binary Programming with Application to
		Capital-Budgeting Problems}.
	\newblock \bibinfo{journal}{\emph{Operations research}} \bibinfo{volume}{18},
	\bibinfo{number}{3} (\bibinfo{year}{1970}), \bibinfo{pages}{454--461}.
	\newblock
	
	
	\bibitem[\protect\citeauthoryear{M{\"u}cke, Piatkowski, and Morik}{M{\"u}cke
		et~al\mbox{.}}{2019}]%
	{muecke2019}
	\bibfield{author}{\bibinfo{person}{Sascha M{\"u}cke}, \bibinfo{person}{Nico
			Piatkowski}, {and} \bibinfo{person}{Katharina Morik}.}
	\bibinfo{year}{2019}\natexlab{}.
	\newblock \showarticletitle{Learning {{Bit}} by {{Bit}}: {{Extracting}} the
		{{Essence}} of {{Machine Learning}}}. In
	\bibinfo{booktitle}{\emph{Proceedings of the {{Conference}} on "{{Lernen}},
			{{Wissen}}, {{Daten}}, {{Analysen}}" ({{LWDA}})}}
	\emph{(\bibinfo{series}{{{CEUR Workshop Proceedings}}})},
	Vol.~\bibinfo{volume}{2454}. \bibinfo{pages}{144--155}.
	\newblock
	
	
	\bibitem[\protect\citeauthoryear{Neukart, Compostella, Seidel, {von Dollen},
		Yarkoni, and Parney}{Neukart et~al\mbox{.}}{2017}]%
	{neukart2017}
	\bibfield{author}{\bibinfo{person}{Florian Neukart}, \bibinfo{person}{Gabriele
			Compostella}, \bibinfo{person}{Christian Seidel}, \bibinfo{person}{David {von
				Dollen}}, \bibinfo{person}{Sheir Yarkoni}, {and} \bibinfo{person}{Bob
			Parney}.} \bibinfo{year}{2017}\natexlab{}.
	\newblock \showarticletitle{Traffic {{Flow Optimization Using}} a {{Quantum
				Annealer}}}.
	\newblock \bibinfo{journal}{\emph{Frontiers in ICT}}  \bibinfo{volume}{4}
	(\bibinfo{year}{2017}).
	\newblock
	
	
	\bibitem[\protect\citeauthoryear{Pardalos and Jha}{Pardalos and Jha}{1992}]%
	{pardalos1992}
	\bibfield{author}{\bibinfo{person}{Panos~M Pardalos} {and}
		\bibinfo{person}{Somesh Jha}.} \bibinfo{year}{1992}\natexlab{}.
	\newblock \showarticletitle{Complexity of Uniqueness and Local Search in
		Quadratic 0\textendash 1 Programming}.
	\newblock \bibinfo{journal}{\emph{Operations research letters}}
	\bibinfo{volume}{11}, \bibinfo{number}{2} (\bibinfo{year}{1992}),
	\bibinfo{pages}{119--123}.
	\newblock
	
	
	\bibitem[\protect\citeauthoryear{Stollenwerk, Lobe, and Jung}{Stollenwerk
		et~al\mbox{.}}{2019}]%
	{stollenwerk2019}
	\bibfield{author}{\bibinfo{person}{Tobias Stollenwerk},
		\bibinfo{person}{Elisabeth Lobe}, {and} \bibinfo{person}{Martin Jung}.}
	\bibinfo{year}{2019}\natexlab{}.
	\newblock \showarticletitle{Flight {{Gate Assignment}} with a {{Quantum
				Annealer}}}. In \bibinfo{booktitle}{\emph{Quantum {{Technology}} and
			{{Optimization Problems}}}} \emph{(\bibinfo{series}{Lecture {{Notes}} in
			{{Computer Science}}})}. \bibinfo{publisher}{{Springer International
			Publishing}}, \bibinfo{pages}{99--110}.
	\newblock
	\urldef\tempurl%
	\url{https://doi.org/10.1007/978-3-030-14082-3_9}
	\showDOI{\tempurl}
	
	
\end{thebibliography}
\end{document}